\documentclass[twocolumn,amsmath,amssymb, showpacs]{revtex4-1}
\usepackage{graphicx}
\usepackage{dcolumn}
\usepackage{bm}
\usepackage{amsthm}
\usepackage{amsmath}
\usepackage{amssymb}
\usepackage{color}

\definecolor{Red}{rgb}{1,0,0}
\def\vec#1{{\bm #1}}

\def\ket#1{| #1 \rangle}

\def\dim{\operatorname{dim}}

\def\Span{\operatorname{span}}

\def\Tr{\operatorname{Tr}}

\def\D{\mathcal{D}}
\def\E{\mathcal{E}}

\def\H{\mathcal{H}}

\def\M{\mathcal{M}}
\def\N{\mathcal{N}}
\def\O{\mathcal{O}}

\def\J{\mathcal{J}}
\def\S{\mathcal{S}}
\def\ONE{\mathbb{I}}

\def\CC{\mathbb{C}}

\def\DFS{\mathop{\rm DFS}}
\def\MNS{\mathop{\rm MNS}}

\def\Fi{\mathop{\rm Fi}}

\begin{document}

\title{Minimal noise subsystems}

\author{Xiaoting Wang$^{1,2,4}$, Mark Byrd$^3$, and Kurt Jacobs$^{1,4,5}$}

\affiliation{ 
$^1$Department of Physics, University of Massachusetts at Boston, Boston, MA 02125, USA \\
$^2$Research Laboratory of Electronics, Massachusetts Institute of Technology, Cambridge, Massachusetts 02139, USA\\
\mbox{$^3$Physics Department \& Computer Science Department, Southern Illinois University, Carbondale, Illinois 62901, USA}\\ 
$^4$Advanced Science Institute, RIKEN, Wako-shi 351-0198, Japan \\ 
$^5$Hearne Institute for Theoretical Physics, Louisiana State University, Baton Rouge, LA 70803, USA
} 


\begin{abstract}
The existence of a decoherence-free subspace/subsystem (DFS) requires that the noise possesses a symmetry. In this work we consider noise models in which perturbations break this symmetry, so that the DFS for the unperturbed model experiences noise. We ask whether in this case there exist subspaces/subsystems that have less noise than the original DFS. We develop a numerical method to search for such \textit{minimal noise subsystems} and apply it to a number of examples. For the examples we examine, we find that if the perturbation is local noise then there is no better subspace/subsystem than the original DFS. We also show that if the noise model remains collective, but is perturbed in a way that breaks the symmetry, then the minimal noise subsystem is distinct from the original DFS, and improves upon it. 
\end{abstract}

\maketitle

\emph{Introduction.}---Error reduction and correction techniques are crucial for realizing scalable and fault-tolerant quantum information processing~\cite{PhysRevA.57.3276}. 
The key technique that enables noise reduction is the encoding of information in a way that includes redundancy. In \emph{quantum error-correction codes} (QECC)~\cite{Shor:95,Calderbank:96,Steane,Gottesman:97,Gottesman:97b,Knill:97b,Gaitan:book} the encoded information is still affected by the noise, but errors can be detected and corrected by exploiting the redundancy. If the noise contains an appropriate symmetry then redundancy can be used to eliminate the noise entirely by encoding in a so-called decoherence free subspace or subsystem (DFS)~\cite{PhysRevLett.79.3306,Lidar98, Lidar99, Zanardi99, Bacon00, Viola00, PhysRevA.63.012301,Wu02,PhysRevLett.91.187903,PhysRevA.72.042303,Bishop-Byrd} assuming that symmetry can be identified. Necessary and sufficient conditions for the existence of a DFS have been derived~\cite{Lidar98,Bacon01}, and numerical methods for finding these DFS structures when exact symmetries exist have been developed~\cite{Holbrook,Knill:06a,Wang13}. Moreover QECC and DFS's can be combined in the implementation of fault-tolerant quantum computation~\cite{Bacon01}. Compared with QECC the implementation of a DFS is simpler and can save computational resources, but is limited to noise that contains one or more symmetries, and this is often absent in real devices. In many cases the real noise can be considered as a (slight) deviation from noise with a symmetry, and the corresponding DFS encoding is still a useful method for noise reduction. Two interesting open questions are the following. For noises that lack a perfect symmetry, can we find a reduced noise subsystem?  Do subsystems exist which experience less noise than the DFS that would be used if the symmetry were perfect? Here we address these questions by developing a numerical method to search for such subspaces or subsystems. We will refer to a subspace or subsystem that experiences the least noise, quantified in some specific way, as a \emph{minimal-noise subspace/subsystem} (MNS). 

\emph{Existence of a DFS.}---All quantum systems $S$ are subject to noise from their environments, and as a result their evolution is not unitary. Under the Born-Markov approximation, the reduced dynamics, excluding the evolution due to the Hamitonian of the system, $H$, is given by $\dot \rho=\sum_i \D[V_i]\rho$, where $\D[V]\rho=V\rho V^\dag -\frac{1}{2}(\rho V^\dag V+  V^\dag V\rho)$. The superoperator $\D[V_i]$ represents the information loss and decoherence in $\S$ and describes the source of noise. The above dynamics is unitary if and only if $\sum_i \D[V_i]\rho=0$, which is equivalent to $[V_i,\rho]=0$ for each $V_i$~\cite{Bacon01}. Thus a DFS is a subspace or a subsystem $\H_0$ such that $[V_i,\rho]=0$ for any $\rho \in \H_0$. Another way of representing a noisy quantum evolution is to use the operator-sum representation, $\E(\rho)=\sum_{k=1}^p E_k \rho E_k^\dag$, where the quantum channel $\E$: $\rho \to \E(\rho)$, is characterized by a set of noise operators $\{E_k\}$ satisfying $\sum_k E_k^\dag E_k=\ONE$. In the following, we assume $H=0$ (i.e., $\S$ functions as a quantum memory) and focus on the effect of the noise. (Alternatively, we can simply transform to the rotating frame.)  An equivalent condition for a DFS is $[E_k, \rho]=0$ for each $k$ and for any $\rho$ in the subspace/subsystem~\cite{Bacon01}. The relation between the noise $\{E_k\}$ and the existence of DFS can be nicely illustrated using the Wedderburn decomposition~\cite{Wedderburn,Barker,Gijswijt}:
\begin{subequations}\label{eqn:alge_decom}
\begin{align}
\N&=\bigoplus_i^\ell \N_i\equiv \bigoplus_i^\ell \ONE_{n_i}\otimes \M_{m_i}, \\
\N'&=\bigoplus_i^\ell \N_i' \equiv\bigoplus_i^\ell \M_{n_i}\otimes \ONE_{m_i}. 
\end{align}
\end{subequations}
Here $\N$ is the $\CC^*$-algebra generated by $\{E_k\}$, and $\N'$ as its commutant algebra. The symbol $\M_{n_i}$ denotes the $n_i\times n_i$ matrix $*$-algebra and $\ONE_{m_i}$ is the identity operator. Hence, any subsystem $\M_{n_i}\otimes \ONE_{m_i}$ with $n_i>1$ corresponds to a DFS that can encode an $n_i$-dimensional quantum system $\rho_{n_i}$ into $\rho^{(en)}=\rho_{n_i}\otimes \frac{1}{m_i}\ONE_{m_i}$.

\emph{Minimal noise subsystems.}---We assume the quantum system is composed of $n_q$ qubits, so that the total dimension of the space is $N = 2^{n_q}$, and we wish to encode the state of a single qubit, which we will denote by $\rho_1$. Inspired by the algebraic structure (\ref{eqn:alge_decom}), we consider the following optimization problem. For a given noisy channel $\E$ with noise operators $E_k$, let $U$ be a unitary matrix that transforms the noise operators from $E_k$ to $UE_kU^\dag$ such that in the new basis, the original state $\rho_1$ is encoded as $\rho=\rho_1\otimes \frac{1}{N_2}\ONE_{N_2}\oplus 0_{N_3}$, with the Hilbert space decomposition $\H=\H_1\otimes \H_2\oplus \H_3$, and $\dim(\H_k)=N_k$ and $N=2^{n_q}=N_1N_2+N_3$ The noise evolution is now  
\vspace{-0.3cm}
\begin{align}\label{eqn:reduced-channel}
\E(\rho)=\sum_{k=1}^p (UE_kU^\dag )  \rho (UE_kU^\dag )^\dag . 
\end{align}
Since the subspace/subsystem encoding scheme is fully characterized by the transformation matrix $U$, we call $U$ the encoding matrix. The reduced evolution on $\H_1$ is 
\vspace{-0.3cm}
\begin{align}\label{eqn:process}
\E_1(\rho_1)\equiv \E|_{\H_1}(\rho) =p_1\rho_1+ \sum_{k=2}^\ell A_k \rho_1 A_k^\dag
\end{align}
where $A_1=\sqrt{p_1} \, \ONE_{N_1}$ and $0<p_1\le 1$. If $\H_1\otimes \H_2$ corresponds to a perfect DFS encoding, then $p_1=1$; if a DFS does not exist, then the optimal $p_1<1$ corresponds to the best encoding scheme, being the MNS. Hence the optimization problem is to find a $U$ that maximizes $p_1$. 
 
Defining $P$ as the projection operator onto the encoded subsystem $\H_1\otimes \H_2$, we have 
\begin{eqnarray}
\E_1(\rho_1) & = & \Tr_2\big( \sum_{k=1}^p (PUE_kU^\dag P) \rho (PU E_k^\dag U^\dag P) \big) \nonumber \\
& = & \frac{1}{N_2}\sum_{k=1}^p \sum_{mm'n} a_{mn}^{(k)} a_{m'n}^{*(k)} \sigma_m^{(1)} \rho_1  \sigma_{m'}^{(1)} 
\end{eqnarray}
where each $PUE_kU^\dag P$ is decomposed into $a_{mn}^{(k)} \sigma_m^{(1)}\sigma_n^{(2)}$. The set of operators $\{\sigma_m^{(j)}\}$ is a generalized orthonormal Pauli basis for Hermitian operators on $\H_j$, $j=1,2$, and $\sigma_0^{(j)}=\frac{1}{\sqrt{N_j}}\ONE_{N_j}$. Thus, the objective function (the function to maximize), as a function of the encoding unitary, $U$, is  
\begin{align}\label{eqn:opti_fn}
\J[U] & \equiv p_1=\frac{1}{N_1N_2}\sum_{k=1}^p \sum_{n} a_{0n}^{(k)} a_{0n}^{*(k)}\nonumber\\
& =\frac{1}{N_1N_2}\sum_{k=1}^p \sum_{n=1}^{N_2^2} |\Tr(PUE_kU^\dag P\sigma_0^{(1)}\sigma_n^{(2)})\nonumber|^2 . 
\end{align}

There are several ways of parametrizing $U$ in terms of $N^2$ real variables, and we adopt the method devised in~\cite{unitary-parametrize} to represent $U$ in terms of $\frac{1}{2}N(N+1)$ phase variables and $\frac{1}{2}N(N-1)$ angle variables: $U(\boldsymbol \alpha)=U(\boldsymbol \alpha^{(phase)},\boldsymbol \alpha^{(angle)})$. Now the optimization of $\J$ is a multivariable optimization problem over the set of variables $\boldsymbol \alpha$. We can use any standard gradient optimization method to find local optimal solution, and we use the BFGS quasi-Newton method~\cite{Nocedal06}.   

More specifically, for each value of $N_1$ and $N_2$, we use the following procedure. First, for the given $\E$ we write down the objective function $J$ in terms of the noise operators $E'_k$, which themselves contain the parameters of the encoding matrix $U=U(\boldsymbol \alpha^{(phase)},\boldsymbol \alpha^{(angle)})$. Then we choose a random initial point $\boldsymbol \alpha^{(0)}$ as the starting point for the numerical search, with $\J^{(0)}=\J[\boldsymbol \alpha^{(0)}]$. The entire optimization process is composed of several iterations. At the $k$-th iteration we use the BFGS quasi-Newton method to update the Hessian and obtain a value of $\J^{(k)}= \J[\boldsymbol \alpha^{(k)}]$, such that $\J^{(k)}>\J^{(k-1)}$. After a sufficient number of iterations, we obtain a sequence of $\{ \J^{(k)} \}$ which monotonically converges to a local maxima $\J_{opt}$, and the encoding matrix $U$ converges to a (locally) optimal encoding matrix $U_{opt}$. We can perform this procedure a number of times with different randomly chosen initial points, and if we continue to obtain the same final value for $\J_{opt}$ we become more sure that the corresponding $U_{opt}$ gives an optimal encoding scheme for the noise $\E$ and the given choices of $N_1$ and $N_2$. Notice that for different values of $N_1$ and $N_2$, $\H_1\otimes \H_2$ represents different subsystems. Hence we need to run the optimization routine for all possible values of $N_1$ and $N_2$, satisfying $N_1N_2\le N$, and derive the MNS in each case. Different choices for $N_1$ and $N_2$ will in general give different values for $\J_{opt}$. The entire algorithm is summarized in Table~\ref{tab:1}.
\begin{table}
\begin{ruledtabular}
\begin{tabular}{ll}
\textbf{Step 1}: & (a) choose $N_1$ and $N_2$ for the encoding subsystem;\\ 
& (b) parametrize $U[\boldsymbol \alpha]=U(\alpha_1,\ldots,\alpha_{N_1^2})$;\\
& (c) express $\J$ in terms of $\boldsymbol \alpha$;\\
\hline
\textbf{Step 2}: &(d) choose a random $\boldsymbol \alpha^{(0)}$ as the initial point;\\
& (e) at the $k$-th iteration, BFGS method gives $\J^{(k)}$;\\
& (f)  $\{ \J^{(k)} \}$ converges to an optimal value $\J_{opt}$;\\
\hline
\textbf{Step 3}: &(g) repeat \textbf{Step 1} and \textbf{Step 2} for other $N_{1}$ and $N_{2}$.
\end{tabular}
\end{ruledtabular}
\caption{Algorithm to search for MNS. \label{tab:1}}
\end{table}

\emph{Applying the procedure to Lindblad evolution}---As mentioned in the introduction, noisy quantum dynamics is often expressed in terms of the Lindblad master equation, rather than Krauss operators, and there is in fact a simple connection between these two representations: Within a short time $dt$, the Lindblad dynamics $\dot{\rho} = \sum_i \D[V_i]\rho$ is equivalent to:
\begin{align}
\E (\rho(0))=\rho(dt)=\sum_k E_k \rho(0) E_k^\dag 
\end{align}\label{eqn:kraus3}
where $E_0 =\ONE- \frac{1}{2} \sum_i V_i^\dag V_i dt$, $E_k = \sqrt{dt} V_k$, $k\ge 1$. Then it is easy to verify that 
\begin{align} \label{eqn:lind_approxi}
\rho(dt)=  \rho(0)-\sum_i \D[V_i]\rho(0) dt+\O(dt)^2
\end{align}  	
We also require $dt> t_1$ where $t_1$ is the characteristic timescale over which the Markovian assumption is valid. As long as $dt$ is small enough, we can always express the given Lindblad dynamics in the Kraus operator-sum form, with $E_k$ as functions of $V_i$. Then we can use the MNS-finding algorithm for the set of noise operators $\{ E_k\}$. 

\emph{Finding a DFS.}---As a test of our algorithm we use it to find the MNS for a system that contains a DFS, since in this case the MNS should coincide with the DFS. Notice that our algorithm requires no prior information of the Wedderburn decomposition (\ref{eqn:alge_decom}), so it is distinct from the previous methods given in~\cite{Holbrook, Wang13}. We chose as our example an $n_q$-qubit system $S_{cn}$ governed by the following dynamics:
\begin{align}\label{eqn:lindblad_colletive}
\dot \rho=\gamma_x\D[S_x] \rho + \gamma_z\D[S_z] \rho
\end{align}
with $S_x = \sum_{k=1}^{n_q} X_k$, $S_z = \sum_{k=1}^{n_q} Z_k$ and decoherence rates $\gamma_{x,z}$. As shown above, we can rewrite the Lindblad evolution of $\rho(dt)$ in the operator-sum representation, where the Kraus operators are 
\begin{align}\label{eqn:Kraus_op_collective_noise}
E_0 = \ONE -\frac{1}{2} (S_x^2+S_y^2)dt, \, E_1 = \sqrt{dt} S_x, \, E_2 = \sqrt{dt} S_z . 
\end{align}
This system has a DFS, and the DFS structure is illustrated through the Wedderburn decomposition~\cite{Bacon01,Byrd:06}. For example, for $n_q=3$, the noise algebra and its commutant are:
\begin{align}\label{eqn:nq3}
\N=\big( \ONE_2\otimes M_2\big) \oplus M_4, \, \N'=\big (M_2\otimes \ONE_2\big )\oplus \ONE_4
\end{align}
We can see that the component $M_2\otimes \ONE_2$ corresponds to a DFS that can store one qubit of information. Now we apply the above algorithm to find the MNS. Let $U$ be encoding matrix, and in new basis, the encoded state is $\rho=\big(\rho_1\otimes \ONE_2/2\big) \oplus 0_4$. The objective function becomes:
\begin{align*}
\J(U)= \frac{1}{8}\sum_{k=1}^3 \sum_{n=1}^{4}|\Tr(PUE_kU^\dag P\sigma_n^{(2)})|^2 ,
\end{align*}
Following the three steps in Table~\ref{tab:1}, we numerically find the optimal encoding matrix $U_{\MNS}$ and find that this coincides with the DFS encoding matrix $U_{\DFS}$. Notice that, for different initial values of $U^{(0)}$, the MNS algorithm may give different $U_{\MNS}$'s. But these all correspond to the same DFS structure $\M_2\otimes \ONE_2$, which is uniquely determined by the Wedderburn decomposition.  

In order to quantify the performance of the encoding matrix $U$, we use the concept of worst-case fidelity after a total time $t_f$, defined to be $\Fi^{wc}[U]=\min_{\rho} \Tr\big(\rho (\mathcal{O}_{\mbox{\scriptsize de}} \cdot\E\cdot\mathcal{O}_{\mbox{\scriptsize en}} )[\rho]\big)$, where $\rho$ is an arbitrary input state, $\mathcal{O}_{\mbox{\scriptsize en}}$ and $\mathcal{O}_{\mbox{\scriptsize de}}$ are the encoding and the decoding actions corresponding to $U$, and $\E$ is the noisy channel at the final time $t_f$. The larger $\Fi_{wc}$ the better the encoding provided by $U$. If $\Fi^{wc}[U]=1$, then $U$ corresponds to a perfect DFS encoding; if $\Fi^{wc}[U_1]>\Fi^{wc}[U_2]$, then the encoding $U_1$ is better than $U_2$. For the collective noise model, we have $\Fi^{wc}[U_{\MNS}]=1$, i.e., the MNS is the same as the DFS. 

\emph{Noise with symmetry-breaking perturbations.}---As mentioned earlier, symmetry is crucial for the existence of a DFS, so when the noise model has no symmetry, the above MNS algorithm is the only way to find the best subspace/subsystem encoding. However, if the noise model is highly asymmetric our results indicate that no MNS's provide significantly reduced noise, and this is not unexpected. Nevertheless, if the noise model can be considered to be a perturbation of a symmetric noise model, we can still use the DFS for the symmetric model to obtain a relatively good encoding scheme. The question is whether there exist MNS's that can provide a better encoding. 

As our first example we consider an $n_q$-qubit system under the collective noise model $S_z$, which applies to trapped-ions~\cite{Winland_ions_decoherence01, PhysRevLett.103.200503}. As the symmetry-breaking perturbation we add local dephasing noise for each qubit and the resulting dynamics is described by 
\begin{align}\label{eqn:collective-Z-perturbed}
\dot \rho= \gamma_z\D[S_z] \rho + \delta \sum_k \gamma_k\D[Z_k] \rho , 
\end{align}
\vspace{-0.3cm}\\
where $\delta$ is the perturbation amplitude, and is chosen to be small. For $n_q=3$, without the local noise terms, the system has two perfect DFS's, generated by $\H_1=\Span\{ \ket{001},\ket{010},\ket{100} \}$ and $\H_2=\Span\{ \ket{101},\ket{110},\ket{011} \}$, and they can be used to encode two independent qutrits. When the local noises are added in, the collective symmetry is broken, and there is no perfect DFS. However, we can apply the MNS algorithm to find the least-noise encoding scheme to encode a qutrit or a qubit. To encode a qutrit, we choose $\N'=(M_3\otimes \ONE_1)\oplus \ONE_5$ in the MNS algorithm. After the optimization routine in Table~\ref{tab:1}, we find $M_3$ is either $\H_1$ or $\H_2$. Similarly, to search for an MNS encoding a qubit, we choose $\N'=(M_2\otimes \ONE_1)\oplus \ONE_6$, and the MNS we find is a 2-D subspace of either $\H_1$ or $\H_2$, depending on the value of $\gamma_k$. For instance, for $\gamma_1=0.33$, $\gamma_2=0.47$, $\gamma_3=0.85$, we find the 2-D MNS is always a subspace of $\H_2$. 

As a second example, we consider the collective noise model in~(\ref{eqn:lindblad_colletive}) perturbed by local noise. Again, we find that the MNS is the same as the DFS for the unperturbed system as long as the perturbation amplitude $\delta$ is sufficiently small. Both examples illustrate that there is no better subspace or subsystem encoding than the original DFS scheme for the collective noise model perturbed by local noise.   
\begin{figure}
\includegraphics[width=\columnwidth]{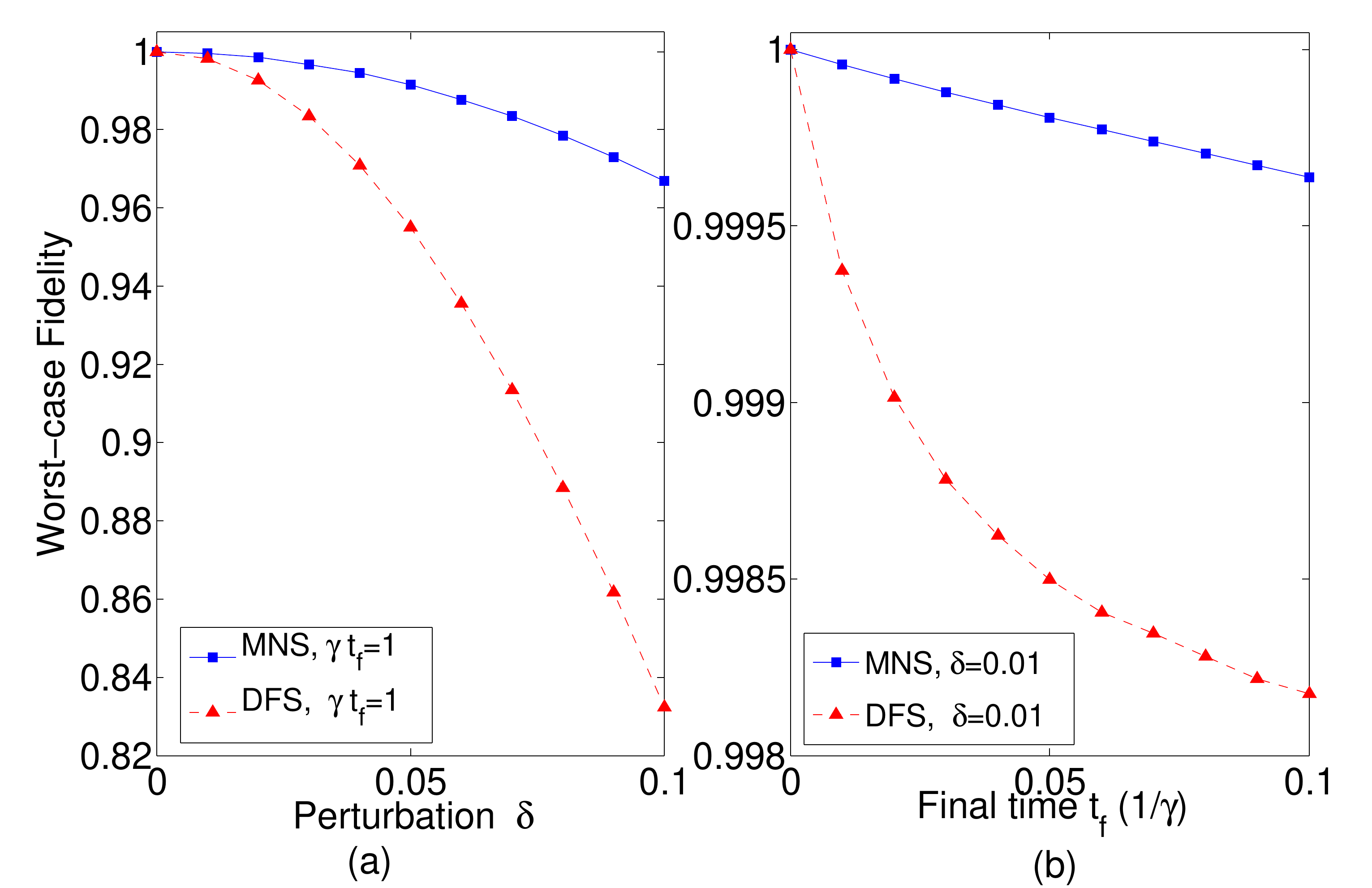}
\caption{(Color online) Worst-case fidelity curves under two different encodings $U_{\MNS}$ and $U_{\DFS}$ for (a) $\delta\in [0,0.1]$ and $\gamma t_f=1$; (b) $\gamma t_f\in [0, 0.1]$ and $\delta=0.1$.}\label{fig:1}
\end{figure}
\begin{figure}
\includegraphics[width=\columnwidth]{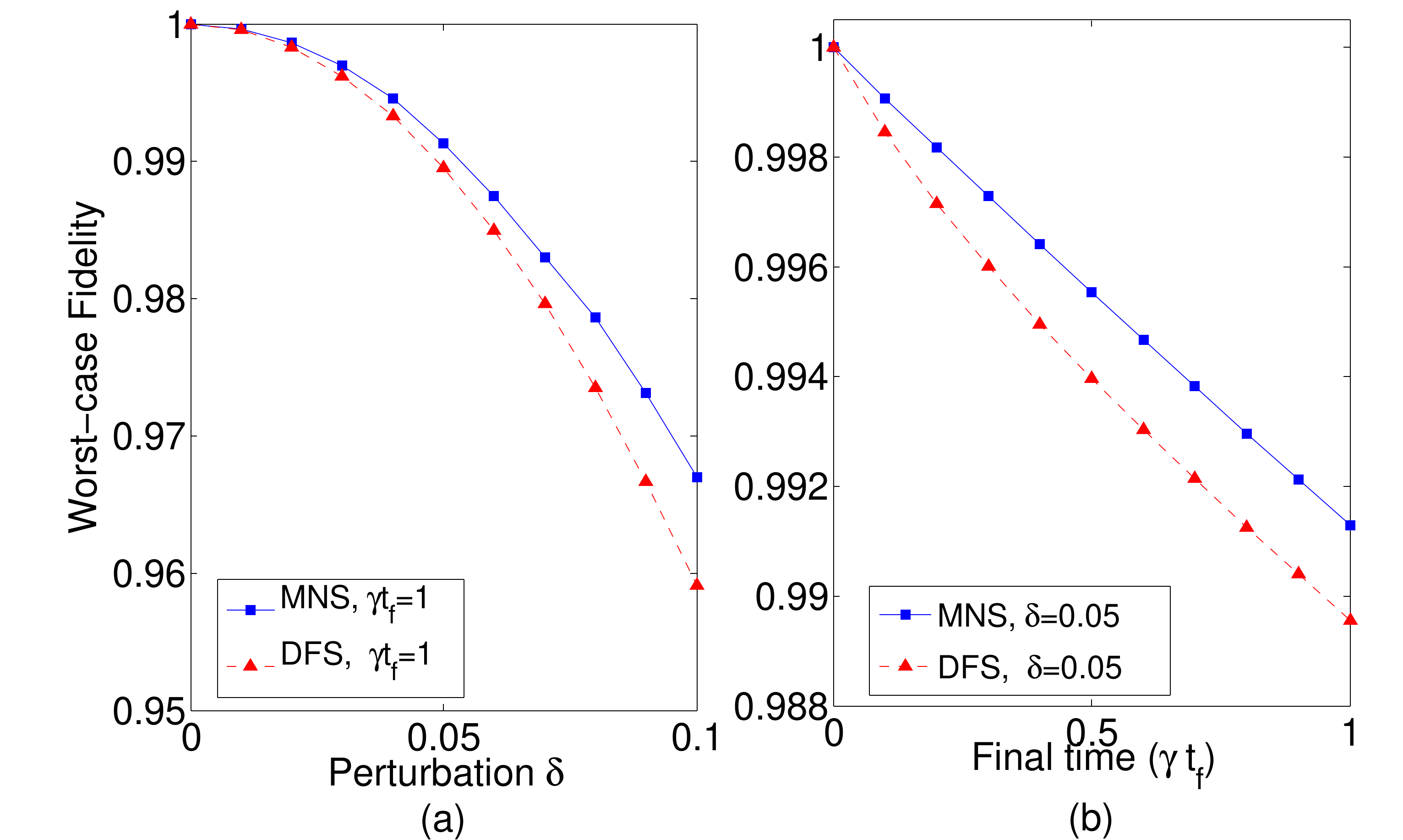}
\caption{(Color online) Worst-case fidelity curves for system perturbed by local random $V_\delta$ under two different encodings $U_{\MNS}$ and $U_{\DFS}$ for (a) $\delta\in [0,0.1]$ and $\gamma t_f=1$; (b) $\gamma t_f\in [0, 0.1]$ and $\delta=0.05$.}\label{fig:2}
\end{figure}

As our third and forth examples we consider a collective noise model in which the collective Lindblad operator is perturbed by i) a randomly chosen global unitary, and ii) a unitary formed by the tensor product of single-qubit, independently selected random unitaries. In this case the noise remains collective, in that there is a single noise channel, but the symmetry is broken so that there is no longer a DFS. We chose as the collective noise model the Lindblad master equation with Linblad operators $S_x$ and $S_z$ as defined above. The noisy dynamics of the perturbed model, $\S$, is given by $\dot\rho=\gamma_1\D[V_\epsilon S_x V_\epsilon^\dag] \rho + \gamma_2\D[S_z] \rho$, where $V_\epsilon$ represents the random global unitary perturbation satisfying $||V_\epsilon-\ONE||=\epsilon$, with $\epsilon$ sufficiently small. The effect of $V_\epsilon$ is to break the collective symmetry since the noise contains independent noise terms, and the result is that no DFS exists. One way of generating such a $V_\epsilon$ is to parameterize $V_{\epsilon}$ using $\frac{1}{2}N(N+1)$ phase variables and $\frac{1}{2}N(N-1)$ angular variables $V_\epsilon(\boldsymbol \beta)=V_\epsilon(\boldsymbol \beta^{(phase)},\boldsymbol \beta^{(angle)})$. For example, we can choose $\boldsymbol \beta^{(phase)}=\vec 0$, and $||\boldsymbol \beta^{(angle)}||=\delta$. Then $\epsilon$ will increase monotonically with $\delta$, and $V_\epsilon=\ONE$ when $\delta=0$. Hence $\delta$ can also be used to quantify the perturbation amplitude. We can choose $\gamma_1=\gamma_2=\gamma$, as the actual values of $\gamma_k$ are not essential for the existence of $MNS$, and for each value of the perturbation amplitude $\delta$ we use the algorithm to find the optimal encoding matrix $U_{\MNS}^\delta$, as well as its worst-case fidelity. For $\delta=0$ we find $\Fi^{wc}_{\MNS}=1$, so that $U_{\MNS}=U_{\DFS}$ for the unperturbed $\S_{cn}$. For nonzero $\delta$ we obtain a $U_{\MNS}^\delta$ that is different from $U_{\DFS}$ and $\Fi^{wc}_{\MNS}$ is strictly larger than $\Fi^{wc}_{\DFS}$. In Fig.~\ref{fig:1} we display and compared the worst-case fidelity curves under the two encoding matrices $U_{\MNS}$ and $U_{\DFS}$ for (a) $\delta\in [0,0.1]$ at a fixed final time $t_f$, and (b) $t_f\in [0,0.1]$ for a fixed value $\delta=0.1$. For $\delta$ near to zero, there is a flat plateau on both curves, which confirms the fact that the DFS encoding is robust against perturbations~\cite{PhysRevA.60.1944}. As $\delta$ increases, $|\Fi^{wc}_{\MNS}-\Fi^{wc}_{\DFS}|$ increases as well, implying that the DFS encoding is no longer as effective when the perturbation becomes large. 

Finally we choose $V_\epsilon$ to be a tensor product of independently selected local random unitaries, and we find similar results to those above, although there is now less difference between the fidelities of the MNS and the DFS for the unperturbed system. In Fig.~\ref{fig:2} we plot the worst-case fidelities for the MNS and DFS.  

\emph{Conclusion.}---The above examples illustrate the ability of our numerical method to find both decoherence-free and minimal-noise subsystems/subspaces given experimental data. It is important to note that this is true even when the symmetry is not exact.  In the examples we have examined, when a collective noise model is perturbed by noise that is local to each subsystem, the minimal-noise subsystem is merely the DFS for the unperturbed system.  However, when a collective noise model is perturbed by a random unitary transformation, while no DFS exists there is a minimal-noise subsystem that is distinct from the DFS for the unperturbed system, providing an improvement over known methods of identification.  

\section*{Acknowledgements} 

KJ is partially supported by the NSF project PHY-1005571, and KJ and XW are partially supported by the NSF project PHY-0902906 and the ARO MURI grant W911NF-11-1-0268. All the authors are partially supported by the Intelligence Advanced Research Projects Activity (IARPA) via Department of Interior National Business Center contract number D11PC20168. The U.S. Government is authorized to reproduce and distribute reprints for Governmental purposes notwithstanding any copyright annotation thereon. Disclaimer: The views and conclusions contained herein are those of the authors and should not be interpreted as necessarily representing the official policies or endorsements, either expressed or implied, of IARPA, DoI/NBC, or the U.S. Government. 

\bibliography{opti_mns_short_v5.bib}

\end{document}